\documentclass[prl, twocolumn, superscriptaddress, nofootnoteinbib,
amsmath,amssymb]{revtex4}

\usepackage{float}
\usepackage{graphicx}

\newcommand{\NVm}{NV$^-$}
\newcommand{\NVo}{NV$^0$}

\begin{document}

\title{Conversion of neutral nitrogen-vacancy centers to negatively-charged nitrogen-vacancy centers through selective oxidation}
\author{K.-M.C. Fu}
\email{kai-mei.fu@hp.com}
\author{C. Santori}
\author{P.E. Barclay}
\author{R.G. Beausoleil}
\affiliation{Information and Quantum System Lab, Hewlett-Packard
Laboratories, 1501 Page Mill Road, MS1123, Palo Alto, California
94304, USA}

\begin{abstract}

The conversion of neutral nitrogen-vacancy centers to negatively charged nitrogen-vacancy centers is demonstrated for centers created by ion implantation and annealing in high-purity diamond.  Conversion occurs with surface exposure to an oxygen atmosphere at 465~$^\circ$C. The spectral properties of the charge-converted centers are investigated.  Charge state control of nitrogen-vacancy centers close to the diamond surface is an important step toward the integration of these centers into devices for quantum information and magnetic sensing applications.

\end{abstract}

\maketitle

The combination of a long ground-state electron coherence time~\cite{ref:balasubramanian2009usc} and the ability to perform optical spin readout in the negatively charged nitrogen-vacancy center (\NVm) in diamond~\cite{ref:gruber1997sco} has motivated proposals to use this center for both quantum information processing (QIP)~\cite{ref:Benjamin2006bgs, ref:Childress2005ftq} and magnetic sensing applications~\cite{ref:taylor2008hsd}.  The best optical and spin properties have been observed in \NVm~centers deep within the diamond lattice, however for both applications it will be highly desirable to use NV centers very close to the diamond surface.  For QIP applications, surface NV centers can be coupled to on-chip waveguides and microcavities~\cite{ref:fu2008cnv, ref:barclay2009cbm}.  For magnetometry applications, NV centers embedded in either diamond nanoparticles or 2D sensing surfaces will combine high magnetic sensitivity with high spatial resolution.

One common technique to create NV centers near a surface is through ion implantation and high temperature annealing~\cite{ref:davies1976oso}.  Since the best spin and optical properties have been observed in high-purity diamond, it would seem advantageous to begin with this material and implant with nitrogen ions.  Upon annealing, the vacancies created during implantation migrate to form NV centers with the implanted nitrogen.  Recent reports, however, indicate that in high purity diamond the preferred charge state of NV centers created very close to the surface (within 200~nm) is the neutral (\NVo) charge state~\cite{ref:santori2009vdn, ref:gaebel2006psn}.  To date, the attractive optical and spin properties observed in the \NVm~center have not been observed in its neutral counterpart.  Thus the first step toward engineering near-surface NVs for QIP and magnetometry is to demonstrate charge state conversion.

A study of the vertical distribution of NV centers created by ion implantation and annealing indicated that the neutral charge state observed near the diamond surface was due to electronic depletion effects consistent with an acceptor layer at the diamond surface~\cite{ref:santori2009vdn}.  In this depletion region, which can extend several microns in high purity diamond ($<$~5~ppb nitrogen), nitrogen donors are ionized and thus cannot donate an electron to the NV center. One possible acceptor candidate is graphitic defects created during the ion implantation~\cite{ref:ristein2000epd}.  Selective oxidation in air of sp$^2$ bonded carbon (graphite or amorphous) over sp$^3$ bonded carbon (diamond) has previously been demonstrated in detonation-synthesized nanodiamond~\cite{ref:osswald2006css}. Here we show that the same technique can be used to convert \NVo~to \NVm~ for centers created 10-75~nm from a single-crystal diamond surface.  Additionally we compare the low temperature optical properties of the near-surface NV centers to those of centers found deep within the diamond matrix.

\begin{table}
\begin{tabular}{|c|c|c|c|}
  \hline
  sample name & energy & dose & depth \\ \hline
  S1 & 10 keV  & 1e10 cm$^{-2}$ & $14\pm5$~nm \\
  S2 & 10 keV & 1e11 cm$^{-2}$ & $14\pm5$~nm \\
  S3 & 50 keV & 1e11 cm$^{-2}$ & $62\pm14$~nm \\
  \hline
\end{tabular}
  \caption{Implantation conditions for the three N-implanted samples studied. Implantation depths are calculated using SRIM software~\cite{ref:zeigler2008sri}}
  \label{table:implant}
  \end{table}
Three high purity commercial diamond samples grown by chemical vapor deposition (E6, CVD electronic grade) were masked with a TEM grid and implanted with nitrogen (CORE Systems). Implantation conditions are given in Table~\ref{table:implant}.  Each sample was next cut into two pieces.  All samples were annealed at 900~$^\circ$C for 1 hour in an Ar/H$_2$ forming gas.  One of the two pieces was reserved as a control sample. Half of the surface of the second piece was masked with a 125~nm thick SiO$_2$ film deposited by e-beam evaporation.  The three half-masked samples were then annealed at 465~$^\circ$C in an oxygen atmosphere in steps of 15, 15, 60, and 60 minutes.

\begin{figure}
  \centering
  \centerline{\scalebox{1}{\includegraphics{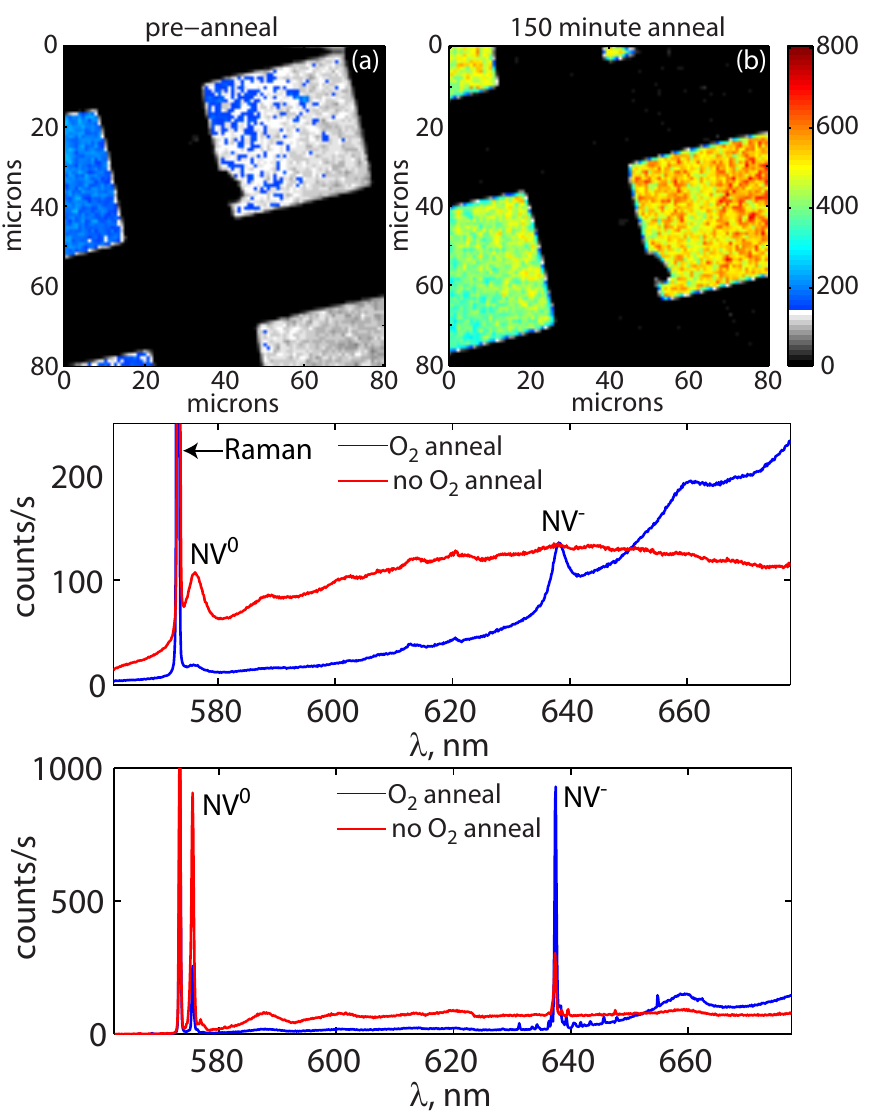}}}
  \caption{a) Confocal image of the unmasked region of sample S2 before annealing.  b) Confocal image of the same region after annealing 150~minutes in O$_2$. Units are counts/ms. c) Room temperature PL spectra of sample S2 in the same implantation region before and after 150 minutes of O$_2$ annealing. 1~mW excitation power. d) 10~K PL spectra of sample S2. 1~mW excitation power.}
  \label{fig:confocal}
\end{figure}

Before, between, and after annealing steps, confocal imaging was used to locate the same implantation region for spectral measurements. An example of a confocal image before and after annealing showing the same implantation square is shown in Figs.~\ref{fig:confocal}a and b.  To obtain these images the sample was excited with a 532~nm laser with 1~mW excitation power focused to a spot with diameter less than 1~$\mu$m while photoluminescence (PL) from the phonon sidebands (650-800~nm) was collected.  After imaging, optical spectra were taken both in the clear and masked regions of each sample for several excitation powers.  Representative spectra of the unmasked region both at room temperature and 10~K before and after annealing are shown in Figs.~\ref{fig:confocal}c and d.  The \NVo~zero phonon line (ZPL) and phonon sidebands are clearly dominant before the O$_2$ anneal.  However after 150 minutes of annealing the centers are predominantly~\NVm.  We note that in most diamond samples both charge states are generally observed in the PL spectra with 532~nm excitation.  This is due to optically induced charge state conversion and is even observed for single NV centers in high-purity diamond~\cite{ref:gaebel2006psn}. The effect of the O$_2$ anneal is to change the relative probability for the NV center to be in the NV$^-$ state.

\begin{figure}
  \centering
  \centerline{\scalebox{1}{\includegraphics{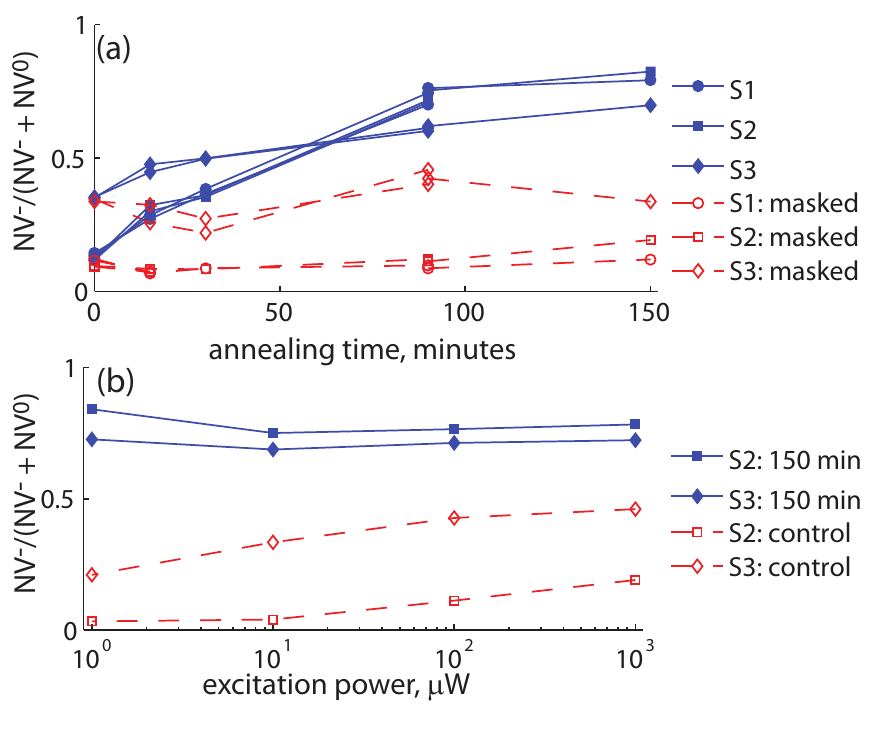}}}
  \caption{a) The room temperature ratio of the \NVm~ZPL intensity to the total ZPL intensity as a function of O$_2$ annealing time. At least two data points were taken on each sample for times $t \leq90$~min. 1~mW excitation power. b) The ratio of the \NVm~ZPL intensity to the total ZPL intensity as a function of optical excitation power at 10~K.}
  \label{fig:graphs}
\end{figure}

The ratio of the room temperature \NVm~ZPL intensity to the total ZPL intensity as a function of annealing time for all three samples is plotted in Fig.~\ref{fig:graphs}a. In all samples the charge state of the NV center switches from predominantly \NVo~to predominantly \NVm~within the first 90 minutes of annealing.  This charge state conversion is not observed in the masked regions of the sample indicating that the oxygen atmosphere is necessary for charge state conversion.  The two 10~keV samples behave similarly, initially exhibiting a very small \NVm~component which rapidly increases.  The 50~keV sample begins with a much larger \NVm~component and a slower conversion is observed.

The excitation power dependence of the \NVm~ZPL component at 10~K for the two implantation energies is plotted in Fig.~\ref{fig:graphs}b. Very little dependence on excitation power is observed in the O$_2$ annealed sample as the excitation power is decreased indicating that the measured ratio is stable and the centers would remain predominantly~\NVm~in the dark.  In contrast, a strong power dependence is observed in both control samples. In these samples the \NVm~component is small at low excitation power and increases with power.  This `photochromatic' effect has been reported previously~\cite{ref:gaebel2006psn} and was attributed to electron excitation and capture dynamics between nearby nitrogen donors and NV centers.  We note that even at the lowest excitation power used, the \NVm~component still appears to be decreasing in the 50~keV (S3) sample. The larger \NVm~component observed in the 50~keV case compared to 10~keV case could be due to a difference in the optically excited electron dynamics for the two implantation depths.

Oxygen annealing at elevated temperatures results in both surface termination with oxygen-containing functional groups and the selective removal of sp$^2$ carbon from diamond samples~\cite{ref:osswald2006css}. Hydrogen surface termination, which is often observed in CVD-grown samples~\cite{ref:landstrass1989rcv}, results in a p-type surface conductive layer~\cite{ref:maier2000osc} which could be the source of an electron depletion layer before the O$_2$ anneal.  This H-terminated surface exhibits both significant surface conductivity ($10^{-4}$-$10^{-5}~\Omega^{-1}$~\cite{ref:maier2000osc}) and a water wetting angle $\theta>80^\circ$~\cite{ref:ostrovskaya2002wse}.  However in both the control and annealed samples, the surface conductivity was less than $10^{-10}~\Omega^{-1}$. Additionally, the water wetting angle before O$_2$ annealing was $\theta = 60^\circ$. Both measurements indicate the absence of a hydrogen-terminated surface.  Upon O$_2$ annealing the water wetting angle decreased to less than $30^\circ$.  This effect has been observed previously in ion-irradiated samples and is attributed to the conversion of irradiation-created sp$^2$ carbon to sp$^3$ carbon~\cite{ref:ostrovskaya2005csw}.  The reduction in wetting angle is due to the adsorption of oxygen functional groups which is assumed to be dependent on the surface bonding states.  These measurements suggest that the mechanism for NV charge-state conversion is linked to the removal of sp$^2$ carbon from the diamond surface.

\begin{figure}
  \centering
  \centerline{\scalebox{1}{\includegraphics{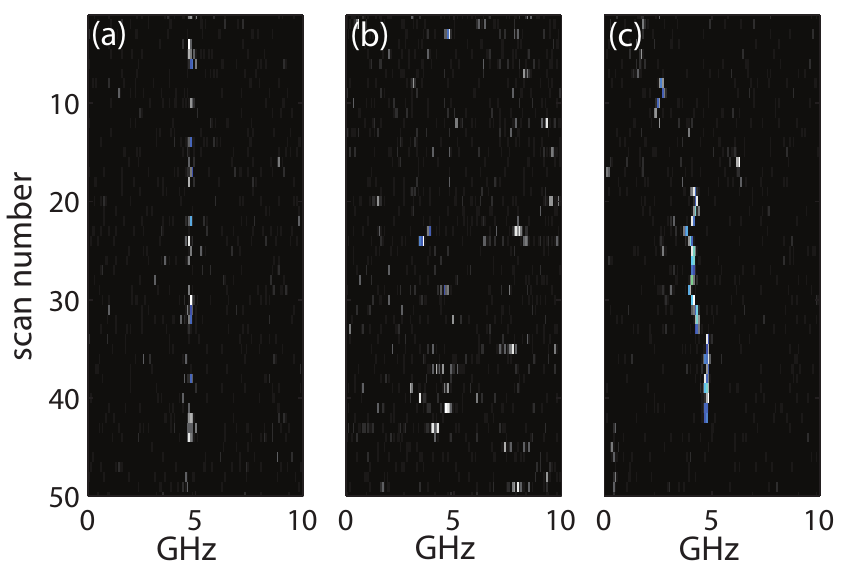}}}
  \caption{ PLE scans of NV centers in sample S1. a) Single NV center formed during the CVD growth process 60~$\mu$m beneath the sample surface.  The repump laser is applied before every scan.  b) NV centers formed by implantation and annealing. The repump laser is applied before every scan.  c) Same as b however the repump laser is applied approximately every 20 scans.}
  \label{fig:ple}
\end{figure}

Once charge state conversion has occurred it is possible to study both the spin and optical properties of the near-surface \NVm~centers.  Here we briefly discuss the low temperature optical properties of the O$_2$ annealed \NVm~centers. For many QIP applications it is desirable for NV centers to emit identical photons~\cite{ref:Benjamin2006bgs, ref:Childress2005ftq}.  Two factors which contribute to the spectral broadening of the emitted photons in the NV system are spectral diffusion and phonon broadening~\cite{ref:fu2009odj}.  High resolution photoluminescence excitation (PLE) spectra of NV centers in sample S1 under different conditions are shown in Figs.~\ref{fig:ple}a-c.  In each case a tunable red diode laser is scanned over the \NVm~ZPL line while PL from the phonon sidebands are detected (650-800~nm).  Measurements were performed at 10~K where phonon broadening of the ZPL is greatly reduced.

In Fig.~\ref{fig:ple}a a single NV which formed during the CVD growth process $\sim60~\mu$m beneath the diamond surface is studied.  Before each scan a green repump laser is applied in order to reverse photo-ionization which eventually occurs with the red laser alone.  During the measurement a single NV track blinks on and off. This blinking is likely caused by charge state conversion~\cite{ref:gaebel2006psn}.  Little spectral diffusion ($< 200$~MHz) is observed between scans which is typical in the highest quality CVD samples.  In Fig.~\ref{fig:ple}b the NV centers formed by implantation and annealing are studied.  Single PL tracks are not observed in this measurement and the NV centers randomly blink on and off throughout the entire scan range.  If the repump laser is applied approximately every 30 scans, as in Fig.~\ref{fig:ple}c, it is possible to observe single NV tracks. However once a particular track goes dark and the repump is applied, the new track may be tens of GHz away from the previous one.  The extreme sensitivity of the \NVm~transition frequency to the repump laser indicates the presence of charge traps close to the NV center.  Future studies, including the effect of surface treatments, will be required to determine if it is possible to reduce the large spectral diffusion observed in these near-surface NV centers.

In summary we have shown that it is possible to convert the charge state of NV centers created by ion implantation and annealing from \NVo~to the desired \NVm~charge state through an oxygen anneal at 465~$^\circ$C.  The successful conversion indicates that the surface depletion layer in high-purity implanted samples is due to graphitic damage which is removed by selective oxidation.  This is an important step toward engineering NV centers near the diamond surface suitable for integration with optical and scanning probe devices for both QIP and magnetometry applications.

This material is based upon work supported by the Defense Advanced Research
Projects Agency under Award No. HR0011-09-1-0006 and The Regents of the University
of California. The authors would like to thank T. Tran for the SiO$_2$ deposition and Y. Gogotsi for helpful discussions.

\bibliography{annealing,qor_qis}

\begin{thebibliography}{18}
\expandafter\ifx\csname natexlab\endcsname\relax\def\natexlab#1{#1}\fi
\expandafter\ifx\csname bibnamefont\endcsname\relax
  \def\bibnamefont#1{#1}\fi
\expandafter\ifx\csname bibfnamefont\endcsname\relax
  \def\bibfnamefont#1{#1}\fi
\expandafter\ifx\csname citenamefont\endcsname\relax
  \def\citenamefont#1{#1}\fi
\expandafter\ifx\csname url\endcsname\relax
  \def\url#1{\texttt{#1}}\fi
\expandafter\ifx\csname urlprefix\endcsname\relax\def\urlprefix{URL }\fi
\providecommand{\bibinfo}[2]{#2}
\providecommand{\eprint}[2][]{\url{#2}}

\bibitem[{\citenamefont{Balasubramanian
  et~al.}(2009)\citenamefont{Balasubramanian, Neumann, Twitchen, Markham,
  Kolesov, Mizuochi, Isoya, Achard, Beck, Tissler
  et~al.}}]{ref:balasubramanian2009usc}
\bibinfo{author}{\bibfnamefont{G.}~\bibnamefont{Balasubramanian}},
  \bibinfo{author}{\bibfnamefont{P.}~\bibnamefont{Neumann}},
  \bibinfo{author}{\bibfnamefont{D.}~\bibnamefont{Twitchen}},
  \bibinfo{author}{\bibfnamefont{M.}~\bibnamefont{Markham}},
  \bibinfo{author}{\bibfnamefont{R.}~\bibnamefont{Kolesov}},
  \bibinfo{author}{\bibfnamefont{N.}~\bibnamefont{Mizuochi}},
  \bibinfo{author}{\bibfnamefont{J.}~\bibnamefont{Isoya}},
  \bibinfo{author}{\bibfnamefont{J.}~\bibnamefont{Achard}},
  \bibinfo{author}{\bibfnamefont{J.}~\bibnamefont{Beck}},
  \bibinfo{author}{\bibfnamefont{J.}~\bibnamefont{Tissler}},
  \bibnamefont{et~al.}, \bibinfo{journal}{Nature Mater.}
  \textbf{\bibinfo{volume}{8}}, \bibinfo{pages}{383} (\bibinfo{year}{2009}).

\bibitem[{\citenamefont{Gruber et~al.}(1997)\citenamefont{Gruber, Drabenstedt,
  Tietz, Fleury, Wrachtrup, and {von Borczyskowski}}}]{ref:gruber1997sco}
\bibinfo{author}{\bibfnamefont{A.}~\bibnamefont{Gruber}},
  \bibinfo{author}{\bibfnamefont{A.}~\bibnamefont{Drabenstedt}},
  \bibinfo{author}{\bibfnamefont{C.}~\bibnamefont{Tietz}},
  \bibinfo{author}{\bibfnamefont{L.}~\bibnamefont{Fleury}},
  \bibinfo{author}{\bibfnamefont{J.}~\bibnamefont{Wrachtrup}},
  \bibnamefont{and} \bibinfo{author}{\bibfnamefont{C.}~\bibnamefont{{von
  Borczyskowski}}}, \bibinfo{journal}{Science} \textbf{\bibinfo{volume}{276}},
  \bibinfo{pages}{2010} (\bibinfo{year}{1997}).

\bibitem[{\citenamefont{Benjamin et~al.}(2006)\citenamefont{Benjamin, Browne,
  Fitzsimons, and Morton}}]{ref:Benjamin2006bgs}
\bibinfo{author}{\bibfnamefont{S.~C.} \bibnamefont{Benjamin}},
  \bibinfo{author}{\bibfnamefont{D.~E.} \bibnamefont{Browne}},
  \bibinfo{author}{\bibfnamefont{J.}~\bibnamefont{Fitzsimons}},
  \bibnamefont{and} \bibinfo{author}{\bibfnamefont{J.~J.~L.}
  \bibnamefont{Morton}}, \bibinfo{journal}{New J. Phys.}
  \textbf{\bibinfo{volume}{8}}, \bibinfo{pages}{141} (\bibinfo{year}{2006}).

\bibitem[{\citenamefont{Childress et~al.}(2005)\citenamefont{Childress, Taylor,
  Sorensen, and Lukin}}]{ref:Childress2005ftq}
\bibinfo{author}{\bibfnamefont{L.}~\bibnamefont{Childress}},
  \bibinfo{author}{\bibfnamefont{J.~M.} \bibnamefont{Taylor}},
  \bibinfo{author}{\bibfnamefont{A.~S.} \bibnamefont{Sorensen}},
  \bibnamefont{and} \bibinfo{author}{\bibfnamefont{M.~D.} \bibnamefont{Lukin}},
  \bibinfo{journal}{Phys. Rev. A} \textbf{\bibinfo{volume}{72}},
  \bibinfo{pages}{52330} (\bibinfo{year}{2005}).

\bibitem[{\citenamefont{Taylor et~al.}(2008)\citenamefont{Taylor, Cappellaro,
  Childress, Jiang, Budker, Hemmer, Yacoby, Walsworth, and
  Lukin}}]{ref:taylor2008hsd}
\bibinfo{author}{\bibfnamefont{J.~M.} \bibnamefont{Taylor}},
  \bibinfo{author}{\bibfnamefont{P.}~\bibnamefont{Cappellaro}},
  \bibinfo{author}{\bibfnamefont{L.}~\bibnamefont{Childress}},
  \bibinfo{author}{\bibfnamefont{L.}~\bibnamefont{Jiang}},
  \bibinfo{author}{\bibfnamefont{D.}~\bibnamefont{Budker}},
  \bibinfo{author}{\bibfnamefont{P.~R.} \bibnamefont{Hemmer}},
  \bibinfo{author}{\bibfnamefont{A.}~\bibnamefont{Yacoby}},
  \bibinfo{author}{\bibfnamefont{R.}~\bibnamefont{Walsworth}},
  \bibnamefont{and} \bibinfo{author}{\bibfnamefont{M.~D.} \bibnamefont{Lukin}},
  \bibinfo{journal}{Nature Phys.} \textbf{\bibinfo{volume}{4}},
  \bibinfo{pages}{810} (\bibinfo{year}{2008}).

\bibitem[{\citenamefont{Fu et~al.}(2008)\citenamefont{Fu, Santori, Barclay,
  Aharonovich, Prawer, Meyer, Holm, and Beausoleil}}]{ref:fu2008cnv}
\bibinfo{author}{\bibfnamefont{K.-M.~C.} \bibnamefont{Fu}},
  \bibinfo{author}{\bibfnamefont{C.}~\bibnamefont{Santori}},
  \bibinfo{author}{\bibfnamefont{P.~E.} \bibnamefont{Barclay}},
  \bibinfo{author}{\bibfnamefont{I.}~\bibnamefont{Aharonovich}},
  \bibinfo{author}{\bibfnamefont{S.}~\bibnamefont{Prawer}},
  \bibinfo{author}{\bibfnamefont{N.}~\bibnamefont{Meyer}},
  \bibinfo{author}{\bibfnamefont{A.~M.} \bibnamefont{Holm}}, \bibnamefont{and}
  \bibinfo{author}{\bibfnamefont{R.~G.} \bibnamefont{Beausoleil}},
  \bibinfo{journal}{Appl. Phys. Lett.} \textbf{\bibinfo{volume}{93}},
  \bibinfo{pages}{234107} (\bibinfo{year}{2008}).

\bibitem[{\citenamefont{Barclay et~al.}(2009)\citenamefont{Barclay, Fu,
  Santori, and Beausoleil}}]{ref:barclay2009cbm}
\bibinfo{author}{\bibfnamefont{P.~E.} \bibnamefont{Barclay}},
  \bibinfo{author}{\bibfnamefont{K.-M.~C.} \bibnamefont{Fu}},
  \bibinfo{author}{\bibfnamefont{C.}~\bibnamefont{Santori}}, \bibnamefont{and}
  \bibinfo{author}{\bibfnamefont{R.~G.} \bibnamefont{Beausoleil}},
  \bibinfo{journal}{Appl. Phys. Lett.} \textbf{\bibinfo{volume}{95}},
  \bibinfo{pages}{191115} (\bibinfo{year}{2009}).

\bibitem[{\citenamefont{Davies and Hamer}(1976)}]{ref:davies1976oso}
\bibinfo{author}{\bibfnamefont{G.}~\bibnamefont{Davies}} \bibnamefont{and}
  \bibinfo{author}{\bibfnamefont{M.~F.} \bibnamefont{Hamer}},
  \bibinfo{journal}{Proc. R. Soc. Lond.~A} \textbf{\bibinfo{volume}{348}},
  \bibinfo{pages}{285} (\bibinfo{year}{1976}).

\bibitem[{\citenamefont{Santori et~al.}(2009)\citenamefont{Santori, Barclay,
  Fu, and Beausoleil}}]{ref:santori2009vdn}
\bibinfo{author}{\bibfnamefont{C.}~\bibnamefont{Santori}},
  \bibinfo{author}{\bibfnamefont{P.~E.} \bibnamefont{Barclay}},
  \bibinfo{author}{\bibfnamefont{K.-M.~C.} \bibnamefont{Fu}}, \bibnamefont{and}
  \bibinfo{author}{\bibfnamefont{R.~G.} \bibnamefont{Beausoleil}},
  \bibinfo{journal}{Phys. Rev. B} \textbf{\bibinfo{volume}{79}},
  \bibinfo{pages}{125313} (\bibinfo{year}{2009}).

\bibitem[{\citenamefont{Gaebel et~al.}(2006)\citenamefont{Gaebel, Domhan,
  Wittmann, Popa, Jelezko, Rabeau, Greentree, Prawer, Trajkov, Hemmer
  et~al.}}]{ref:gaebel2006psn}
\bibinfo{author}{\bibfnamefont{T.}~\bibnamefont{Gaebel}},
  \bibinfo{author}{\bibfnamefont{M.}~\bibnamefont{Domhan}},
  \bibinfo{author}{\bibfnamefont{C.}~\bibnamefont{Wittmann}},
  \bibinfo{author}{\bibfnamefont{I.}~\bibnamefont{Popa}},
  \bibinfo{author}{\bibfnamefont{F.}~\bibnamefont{Jelezko}},
  \bibinfo{author}{\bibfnamefont{J.}~\bibnamefont{Rabeau}},
  \bibinfo{author}{\bibfnamefont{A.}~\bibnamefont{Greentree}},
  \bibinfo{author}{\bibfnamefont{S.}~\bibnamefont{Prawer}},
  \bibinfo{author}{\bibfnamefont{E.}~\bibnamefont{Trajkov}},
  \bibinfo{author}{\bibfnamefont{P.}~\bibnamefont{Hemmer}},
  \bibnamefont{et~al.}, \bibinfo{journal}{Appl. Phys.~B}
  \textbf{\bibinfo{volume}{82}}, \bibinfo{pages}{243} (\bibinfo{year}{2006}).

\bibitem[{\citenamefont{Ristein}(2000)}]{ref:ristein2000epd}
\bibinfo{author}{\bibfnamefont{J.}~\bibnamefont{Ristein}},
  \bibinfo{journal}{Diamond \& Rel. Mat.} \textbf{\bibinfo{volume}{9}},
  \bibinfo{pages}{1129} (\bibinfo{year}{2000}).

\bibitem[{\citenamefont{Osswald et~al.}(2006)\citenamefont{Osswald, Yushin,
  Mochalin, Kucheyev, and Gogotsi}}]{ref:osswald2006css}
\bibinfo{author}{\bibfnamefont{S.}~\bibnamefont{Osswald}},
  \bibinfo{author}{\bibfnamefont{G.}~\bibnamefont{Yushin}},
  \bibinfo{author}{\bibfnamefont{V.}~\bibnamefont{Mochalin}},
  \bibinfo{author}{\bibfnamefont{S.~O.} \bibnamefont{Kucheyev}},
  \bibnamefont{and} \bibinfo{author}{\bibfnamefont{Y.}~\bibnamefont{Gogotsi}},
  \bibinfo{journal}{J. Am. Chem. Soc.} \textbf{\bibinfo{volume}{128}},
  \bibinfo{pages}{11635} (\bibinfo{year}{2006}).

\bibitem[{\citenamefont{Zeigler}(2008)}]{ref:zeigler2008sri}
\bibinfo{author}{\bibfnamefont{J.}~\bibnamefont{Zeigler}},
  \emph{\bibinfo{title}{The stopping range of ions in matter, {SRIM-2008}}}
  (\bibinfo{year}{2008}).

\bibitem[{\citenamefont{Landstrass and Ravi}(1989)}]{ref:landstrass1989rcv}
\bibinfo{author}{\bibfnamefont{M.~I.} \bibnamefont{Landstrass}}
  \bibnamefont{and} \bibinfo{author}{\bibfnamefont{K.~V.} \bibnamefont{Ravi}},
  \bibinfo{journal}{Appl. Phys. Lett.} \textbf{\bibinfo{volume}{55}},
  \bibinfo{pages}{975} (\bibinfo{year}{1989}).

\bibitem[{\citenamefont{Maier et~al.}(2000)\citenamefont{Maier, Riedel, Mantel,
  Ristein, and Ley}}]{ref:maier2000osc}
\bibinfo{author}{\bibfnamefont{F.}~\bibnamefont{Maier}},
  \bibinfo{author}{\bibfnamefont{M.}~\bibnamefont{Riedel}},
  \bibinfo{author}{\bibfnamefont{B.}~\bibnamefont{Mantel}},
  \bibinfo{author}{\bibfnamefont{J.}~\bibnamefont{Ristein}}, \bibnamefont{and}
  \bibinfo{author}{\bibfnamefont{L.}~\bibnamefont{Ley}},
  \bibinfo{journal}{Phys. Rev. Lett.} \textbf{\bibinfo{volume}{85}},
  \bibinfo{pages}{3472} (\bibinfo{year}{2000}).

\bibitem[{\citenamefont{Ostrovskayaa et~al.}(2002)\citenamefont{Ostrovskayaa,
  Perevertailoa, Ralchenkob, Dementjevc, and
  Loginova}}]{ref:ostrovskaya2002wse}
\bibinfo{author}{\bibfnamefont{L.}~\bibnamefont{Ostrovskayaa}},
  \bibinfo{author}{\bibfnamefont{V.}~\bibnamefont{Perevertailoa}},
  \bibinfo{author}{\bibfnamefont{V.}~\bibnamefont{Ralchenkob}},
  \bibinfo{author}{\bibfnamefont{A.}~\bibnamefont{Dementjevc}},
  \bibnamefont{and} \bibinfo{author}{\bibfnamefont{O.}~\bibnamefont{Loginova}},
  \bibinfo{journal}{Diam. Relat. Mater.} \textbf{\bibinfo{volume}{11}},
  \bibinfo{pages}{845} (\bibinfo{year}{2002}).

\bibitem[{\citenamefont{Ostrovskaya et~al.}(2005)\citenamefont{Ostrovskaya,
  Dementiev, Kulakova, and Ralchenko}}]{ref:ostrovskaya2005csw}
\bibinfo{author}{\bibfnamefont{L.}~\bibnamefont{Ostrovskaya}},
  \bibinfo{author}{\bibfnamefont{A.}~\bibnamefont{Dementiev}},
  \bibinfo{author}{\bibfnamefont{I.}~\bibnamefont{Kulakova}}, \bibnamefont{and}
  \bibinfo{author}{\bibfnamefont{V.}~\bibnamefont{Ralchenko}},
  \bibinfo{journal}{Diam. Relat. Mater.} \textbf{\bibinfo{volume}{14}},
  \bibinfo{pages}{486} (\bibinfo{year}{2005}).

\bibitem[{\citenamefont{Fu et~al.}(2009)\citenamefont{Fu, Santori, Barclay,
  Rogers, Manson, and Beausoleil}}]{ref:fu2009odj}
\bibinfo{author}{\bibfnamefont{K.-M.~C.} \bibnamefont{Fu}},
  \bibinfo{author}{\bibfnamefont{C.}~\bibnamefont{Santori}},
  \bibinfo{author}{\bibfnamefont{P.~E.} \bibnamefont{Barclay}},
  \bibinfo{author}{\bibfnamefont{L.~J.} \bibnamefont{Rogers}},
  \bibinfo{author}{\bibfnamefont{N.~B.} \bibnamefont{Manson}},
  \bibnamefont{and} \bibinfo{author}{\bibfnamefont{R.~G.}
  \bibnamefont{Beausoleil}}, \bibinfo{journal}{Phys. Rev. Lett.}
  \textbf{\bibinfo{volume}{103}}, \bibinfo{pages}{256404}
  (\bibinfo{year}{2009}).

\end{thebibliography}
\bibliographystyle{apsrev}

\end{document}